\documentclass[a4paper,UKenglish,cleveref, autoref, thm-restate]{lipics-v2021}

\newcommand{\ie}{\textit{i.e.}, }
\newcommand{\eg}{\textit{e.g.}, }

\newcommand{\ra}{\mathop{\rightarrow}}
\newcommand{\rra}{\mathop{\Rightarrow}}
\newcommand{\lpa}{\mathop{\leadsto}}

\newcommand{\nat}{\mathbb{N}}
\newcommand{\termset}{T(\Sigma, X)}
\newcommand{\subsset}{S(\Sigma, X)}
\newcommand{\seqset}[1]{{\overline{{#1}}}}

\newcommand{\vars}{\operatorname{\mathit{Var}}}
\newcommand{\dom}{\operatorname{\mathit{Dom}}}
\newcommand{\mgu}{\operatorname{\mathit{mgu}}}

\newcommand{\fsym}{\operatorname{\mathsf{f}}}
\newcommand{\gsym}{\operatorname{\mathsf{g}}}
\newcommand{\ssym}{\operatorname{\mathsf{s}}}
\newcommand{\zero}{\mathsf{0}}

\newcommand{\nti}{\textsf{NTI}}

\title{Binary Non-Termination in Term Rewriting and Logic Programming}
\titlerunning{Binary Non-Termination}

\author{\'Etienne Payet}{LIM - Université de la Réunion,
  France \and \url{http://lim.univ-reunion.fr/staff/epayet/}}
  {etienne.payet@univ-reunion.fr}{https://orcid.org/0000-0002-3519-025X}{}
\authorrunning{\'E. Payet}

\Copyright{Étienne Payet}



\nolinenumbers
\hideLIPIcs
\begin{document}

\maketitle

\begin{abstract}
  We present a new syntactic criterion for the automatic detection
  of non-termination in an abstract setting that encompasses
  a simplified form of term rewriting and logic programming.
\end{abstract}

\section{Introduction}
This paper is concerned with non-termination in structures where
one rewrites elements using indexed binary relations. Such
structures can be formalised by \emph{abstract reduction systems}
(ARSs)~\cite{baaderN98}, \ie couples $(A,\rra_I)$ where
$A$ is a set and $\rra_I$ (the rewrite relation) is
the union of binary relations on $A$, indexed by a set $I$,
\ie $\rra_I = \bigcup \{\rra_{\iota} \mid \iota \in I\}$.
Non-termination in these structures can be formalised as the
existence of an infinite rewrite sequence 
$a_0 \rra_{\iota_0} a_1 \rra_{\iota_1} \cdots$.
\emph{Term rewrite systems (TRSs)} and
\emph{logic programs (LPs)} are concrete instances of ARSs:
$A$ is the set of finite terms and $I$ indicates what rule
(= a couple of finite terms) is applied at what position.
A crucial difference is that the rewrite relation of TRSs
relies on instantiation while that of LPs relies on narrowing,
\ie on unification.
In this paper, we present a new syntactic criterion for the
automatic detection of non-termination in an abstract setting
that encompasses a simplified form of term rewriting and
logic programming. Namely, we suppose that the rewriting
always takes place at the root position of terms
(see Def.~\ref{def:trs-lp-rew-rel} below). There exist
program transformation techniques that make it possible to
place oneself in such a context, \eg the overlap
closure~\cite{guttagKM83} in term rewriting or the binary
unfoldings~\cite{codishT99,gabbrielliG94} in logic
programming preserve the non-termination property of the
original program.

\section{Preliminaries}
We let $\nat$ denote the set of non-negative integers.

\subsection{Binary Relations}
If $\rra$ and $\hookrightarrow$ are binary relations on a set
$A$, then $\rra \mathop{\circ} \hookrightarrow$ denotes their
\emph{composition}. We let $\rra^0$ be the identity relation
and, for all $n\in\nat$, $\rra^{n+1}=(\rra^n \circ \rra)$.
Moreover, $\rra^* = \bigcup \{\rra^n \mid n \geq 0 \}$ is the
\emph{reflexive and transitive closure} of $\rra$. We
formalise non-termination as the existence of an infinite
sequence of connected elements:
\begin{definition}\label{def:chain}
  Let $\rra$ be a binary relation on a set $A$.
  A \emph{$\rra$-chain} is a (possibly infinite) sequence
  $a_0, a_1, \dots$ of elements of $A$ such that
  $a_n \rra a_{n+1}$ for all $n \in \nat$.
  We simply write it as $a_0 \rra a_1 \rra \cdots$. 
\end{definition}

\subsection{Terms}
We use the same definitions and notations as~\cite{baaderN98}
for terms.
From now on, we fix a \emph{signature} $\Sigma$ (the
\emph{function symbols}) together with an infinite countable
set $X$ of \emph{variables}, with $\Sigma \cap X = \emptyset$.
We let $\fsym,\gsym,\ssym$ be function symbols of positive
arity and $\zero$ be a constant symbol.
The set of all \emph{terms} built from $\Sigma$ and $X$ is
denoted by $\termset$.
A \emph{context} is a term with at least one ``hole'',
represented by $\square$, in it. For all terms or contexts $t$,
we let $\vars(t)$ denote the set of variables occurring in $t$
and, for all contexts $c$, we let $c[t]$ denote the term or
context obtained from $c$ by replacing all the occurrences of
$\square$ by $t$. For all contexts $c$, we let
$c^0 = \square$ and, for all $n\in\nat$, $c^{n+1}=c[c^n]$.
Terms are generally denoted by $a,s,t,u,v$, variables by $x,y$
and contexts by $c$, possibly with subscripts and quotes. 

The set $\subsset$ of all \emph{substitutions} consists
of the functions $\theta$ from $X$ to $\termset$ such that
$\dom(\theta)=\{x \in X \mid \theta(x) \neq x\}$ is finite.
A substitution $\theta$ is usually written as
$\{x\mapsto\theta(x) \mid x \in \dom(\theta)\}$ and its
application to a term $s$ as $s\theta$.
A \emph{renaming} is a substitution that is a bijection on $X$.
The \emph{composition} of substitutions $\sigma$ and
$\theta$ is denoted as $\sigma\theta$.
We say that $\sigma$ is \emph{more general than} $\theta$ if
$\theta=\sigma\eta$ for some substitution $\eta$.
We let $\theta^0=\emptyset$
(the identity substitution) and, for all $n\in\nat$,
$\theta^{n+1}=\theta^n\theta$.

A term $s$ is an \emph{instance} of a term $t$ if $s=t\theta$
for some $\theta\in \subsset$. On the other hand, $s$ \emph{unifies}
with $t$ if $s\theta=t\theta$ for some $\theta\in \subsset$;
then, $\theta$ is called a \emph{unifier} of $s$ and $t$ and
$\mgu(s,t)$ denotes the \emph{most general unifier} (mgu) of
$s$ and $t$.

\subsection{Term Rewriting and Logic Programming}
\label{sect:trs-lp}
We refer to~\cite{baaderN98} (resp.~\cite{apt97})
for the basics of term rewriting (resp. logic programming).

\begin{definition}\label{def:prog}
  A \emph{program} is a subset of $\termset^2$,
  every element $(u,v)$ of which is called a \emph{rule},
  where $u$ (resp. $v$) is the \emph{left-hand side} 
  (resp. \emph{right-hand side}).
  For each program $P$, we let $\seqset{P}$ denote the set
  of all finite, non-empty, sequences of elements of $P$.
\end{definition}

In this paper, we only consider ARSs $(A,\rra_I)$ such that 
$A = \termset$ and $I$ is a program. Hence the following
simplified definition.
\begin{definition}
  An \emph{abstract reduction system (ARS)} is a union of
  binary relations on $\termset$ indexed by a program,
  \ie it has the form
  $\rra_P = \bigcup \{\rra_r \subseteq \termset^2 \mid r \in P\}$
  for some program $P$.
  For each ARS $\rra_P$ and each $\omega = (r_1,\dots,r_n)$ in
  $\seqset{P}$, we let
  $\rra_{\omega} = (\rra_{r_1} \circ \cdots \circ \rra_{r_n})$.
\end{definition}

The next definition introduces term rewrite systems and logic
programs as concrete instances of ARSs.
For all terms $s$ and rules $(u,v)$ and $(u',v')$, we write
$(u,v) \ll_s (u',v')$ to denote that $(u, v)$ is a \emph{variant}
of $(u',v')$ \emph{variable disjoint} with $s$, \ie for some
renaming $\gamma$, we have $u = u'\gamma$, $v = v'\gamma$ and
$\vars(u) \cap \vars(s) = \vars(v) \cap \vars(s) = \emptyset$.
\begin{definition}\label{def:trs-lp-rew-rel}
  For each program $P$, we let
  $\ra_P = \bigcup \{\ra_r \mid r\in P \}$
  and
  $\lpa_P = \bigcup \{\lpa_r \mid r\in P \}$
  where, for all $r \in P$,
  \begin{align*}
    \ra_r &= \left\{ \big(u\theta,v\theta\big) \in \termset^2
    \;\middle\vert\;
    (u, v) = r,\ \theta \in \subsset\right\}
    & \text{(Term Rewriting)} \\
    \lpa_r &= \left\{ \big(s,v\theta\big) \in \termset^2
    \;\middle\vert\;
      (u, v) \ll_s r, \ \theta = \mgu(s, u)\right\}
    & \text{(Logic Programming)}
  \end{align*}
  We say that $\ra_P$ (resp. $\lpa_P$) is a
  \emph{term rewrite system} (resp. a \emph{logic program}).
\end{definition}

\begin{example}\label{ex:prel-trs}
  Let $r = \big(\fsym(x), \ssym(x)\big) = (u,v)$.
  Then, $\fsym^2(x) \ra_r \ssym(\fsym(x))$ because
  $\fsym^2(x) = u\theta$ and
  $\ssym(\fsym(x)) = v\theta$ for
  $\theta = \{x\mapsto\fsym(x)\}$.
  Let $r' = \big(\fsym(\gsym(x,\zero)), \fsym(x)\big)$
  and $s = \fsym(\gsym(x,x))$.
  The rule $(u',v') = (\fsym(\gsym(x',\zero)), \fsym(x'))$
  is a variant of $r'$ variable disjoint with $s$. 
  Let $\theta' = \{x\mapsto\zero, x'\mapsto\zero\}$.
  Then, $\theta' = \mgu\left(s, u'\right)$ and we have
  $s \lpa_{r'} v'\theta'$, \ie
  $\fsym(\gsym(x,x)) \lpa_{r'} \fsym(\zero)$.
\end{example}

In term rewriting and in logic programming (modulo a
condition), the left-hand side of a rule can be rewritten
to the corresponding instance of the right-hand side.
\begin{lemma}\label{lem:stability-rules}
  Let $r = (u,v)$ be a rule and $\theta$ be a substitution.
  We have $u\theta \ra_r v\theta$ and, if
  $\vars(v) \subseteq \vars(u)$, $u\theta \lpa_r v\theta$.
\end{lemma}

\section{Binary Non-Termination}\label{sect:bin-nonterm}
%
We are interested in \emph{binary chains}, \ie infinite
chains that consist of the repetition of two sequences
of rules. There are ARSs that admit such chains but no
infinite chain consisting of the repetition of a single
sequence (see, \eg $\ra_P$ in Ex.~\ref{ex:binchain-trs}
and Ex.~\ref{ex:binchain-lp} below).
More precisely:
\begin{definition}\label{def:binary-chain}
  Let $\rra_P$ be an ARS and $\omega_1,\omega_2 \in \seqset{P}$.
  A \emph{$(\omega_1,\omega_2,\rra_P)$-chain} is an
  infinite $(\rra^*_{\omega_1} \circ \rra_{\omega_2})$-chain.
\end{definition}

\begin{example}\label{ex:binchain-trs}
  Let $\rra_P \in \{\ra_P,\lpa_P\}$ where $P$ is the program
  that consists of the rules
  \[r_1 = \big(\fsym(x, \ssym(y)), \fsym(\ssym^2(x), y)\big)
  \qquad
  r_2 = \big(\fsym(x, \zero), \fsym(\ssym(\zero), x)\big)\]
  (see~\cite{zantema96} and \verb+TRS_Standard/Zantema_15/ex12.xml+
  in~\cite{tpdb}). We have the $(r_1,r_2,\rra_P)$-chain:
  \[\fsym(\ssym(\zero), \zero) \rra^0_{r_1}
  \fsym(\ssym(\zero), \zero) \rra_{r_2}
  \fsym(\ssym(\zero), \ssym(\zero)) \rra^1_{r_1}
  \fsym(\ssym^3(\zero), \zero) \rra_{r_2}
  \fsym(\ssym(\zero), \ssym^3(\zero)) \rra^3_{r_1}
  \cdots\]
\end{example}

\begin{example}\label{ex:binchain-lp}
  Let $\rra_P \in \{\ra_P,\lpa_P\}$ where $P$ is the program
  that consists of the rules
  \[r_1 = \big(\fsym(x, \ssym(y)), \fsym(\ssym(x), y)\big)
  \qquad
  r_2= \big(\fsym(x, \zero), \fsym(x, \ssym(x))\big)\]
  (see~\cite{zantema96} and \verb+TRS_Standard/Zantema_15/ex14.xml+
  in~\cite{tpdb}). We have the $(r_1,r_2,\rra_P)$-chain:
  \[\fsym(\zero, \ssym(\zero)) \rra^1_{r_1}
  \fsym(\ssym(\zero), \zero) \rra_{r_2} 
  \fsym(\ssym(\zero), \ssym^2(\zero)) \rra^2_{r_1}
  \fsym(\ssym^3(\zero), \zero) \rra_{r_2}
  \fsym(\ssym^3(\zero), \ssym^4(\zero)) \rra^4_{r_1}
  \cdots\]
\end{example}

Now, we present a criterion for the detection of binary
chains. It is tailored to deal with specific sequences
$\omega_1$ and $\omega_2$ that each consist of a single
rule of a particular form. Intuitively, the rule $r_1$
of $\omega_1$ and the rule $r_2$ of $\omega_2$ are
mutually recursive; in $r_1$, a context $c$ is removed from
the left-hand side to the right-hand side while, in $r_2$,
$c$ is added again. Ex.~\ref{ex:binchain-trs}
and Ex.~\ref{ex:binchain-lp} are concrete instances, with
$c = \ssym(\square)$. This is formalised as follows.
\begin{definition}\label{def:rec-pair}
  A \emph{recurrent pair} for a program $P$ is a
  pair $(r_1,r_2) \in P^2$ such that
  \begin{itemize}
    \item $r_1 = \big( \fsym(x, c[y]), \fsym(c^{n_1}[x], y) \big)$
    and
    $r_2 = \big( \fsym(x, s), \fsym(c^{n_2}[t], c^{n_3}[x]) \big)$
    \item $x \neq y$
    \item $\vars(c) = \vars(s) = \emptyset$
    \item $t \in \{x,s\}$
  \end{itemize}
\end{definition}

\begin{example}
  In Ex.~\ref{ex:binchain-trs}, we have
  $(n_1,n_2,n_3) = (2,1,0)$, $c = \ssym(\square)$
  and $s = t = \zero$.
  In Ex.~\ref{ex:binchain-lp}, we have
  $(n_1,n_2,n_3) = (1,0,1)$, $c = \ssym(\square)$,
  $s = \zero$ and $t = x$.
\end{example}

We show that the existence of a recurrent pair leads to that
of a binary chain (see Prop.~\ref{prop:binary-chain}), provided
that property~\eqref{eq:stability-rules} below is satisfied.
The rest of this section is parametric in an ARS $\rra_P$
and a recurrent pair $(r_1,r_2)$ for $P$ as in
Def.~\ref{def:rec-pair}, with $r_1 = (u_1,v_1)$ and
$r_2 = (u_2,v_2)$. We suppose that we have
\begin{equation}\label{eq:stability-rules}
  \forall \theta \in \subsset \
  (u_1\theta \rra_{r_1} v_1\theta) \land
  (u_2\theta \rra_{r_2} v_2\theta)
\end{equation}
As $\vars(v_1) \subseteq \vars(u_1)$ and
$\vars(v_2) \subseteq \vars(u_2)$, by Lem.~\ref{lem:stability-rules}
both $\ra_p$ and $\lpa_P$ satisfy~\eqref{eq:stability-rules}.

For the sake of readability, we introduce the following
notation.
\begin{definition}
  For all $m, n \in \nat$, we let $\fsym(m, n)$ denote
  the term $\fsym(c^m[s],c^n[s])$. 
\end{definition}

Then, we have the following two lemmas.
Lem.~\ref{lem:reductions-r1} states that $r_1$
allows one to iteratively move a tower of $c$'s
from the second to the first argument of $\fsym$.
Conversely, Lem.~\ref{lem:reductions-r2} states that
$r_2$ allows one to copy a tower of $c$'s from the
first to the second argument of $\fsym$ in just
one step.
\begin{lemma}\label{lem:reductions-r1}
  For all $m,n \in \nat$,
  $\fsym(m, n) \rra^n_{r_1} \fsym(n_1 \times n + m, 0)$.
\end{lemma}
\begin{proof}
  We proceed by induction on $n$.
  \begin{itemize}
    \item (Base: $n = 0$) Here, $\rra^n_{r_1}$ is the identity.
    Hence, for all $m \in \nat$, we have
    $\fsym(m, n) \rra^n_{r_1} \fsym(m, n)$,
    where $\fsym(m, n) = \fsym(n_1 \times n + m, 0)$.
    \item (Induction) Suppose that for some $n \in \nat$
    we have $\fsym(m, n) \rra^n_{r_1} \fsym(n_1 \times n + m, 0)$
    for all $m \in \nat$. Let $m \in \nat$. Then,
    $\fsym(m, n+1) = \fsym(c^m[s], c^{n+1}[s])
    = u_1\{x \mapsto c^m[s], y \mapsto c^n[s]\}$.
    Therefore, by~\eqref{eq:stability-rules}, we have
    $\fsym(m, n+1) \rra_{r_1}
    v_1\{x \mapsto c^m[s], y \mapsto c^n[s]\}$
    where $v_1\{x \mapsto c^m[s], y \mapsto c^n[s]\} =
    \fsym(c^{n_1 + m}[s],c^n[s]) = \fsym(n_1 + m, n)$.
    But, by induction hypothesis, we have
    $\fsym(n_1 + m, n) \rra^n_{r_1}
    \fsym(n_1 \times n + (n_1 + m), 0)$, \ie
    $\fsym(n_1 + m, n) \rra^n_{r_1}
    \fsym(n_1 \times (n + 1) + m, 0)$. Finally,
    $\fsym(m, n+1) \rra^{n+1}_{r_1}
    \fsym(n_1 \times (n + 1) + m, 0)$.
  \end{itemize}
\end{proof}

\begin{lemma}\label{lem:reductions-r2}
  For all $m \in \nat$,
  $\fsym(m, 0) \rra_{r_2} \fsym(m' + n_2, m + n_3)$
  where $m' = 0$ if $t = s$ and $m' = m$ if $t = x$.
\end{lemma}
\begin{proof}
  Let $m \in \nat$. We have
  $\fsym(m, 0) = \fsym(c^m[s], s) = u_2\{x \mapsto c^m[s]\}$.
  Hence, by~\eqref{eq:stability-rules}, we have
  $\fsym(m, 0) \rra_{r_2} v_2\{x \mapsto c^m[s]\}$.
  \begin{itemize}
    \item If $t = s$ then $v_2\{x \mapsto c^m[s]\} =
    \fsym(c^{n_2}[s],c^{m+n_3}[s]) = \fsym(n_2, m+n_3)$.
    \item If $t = x$ then $v_2\{x \mapsto c^m[s]\} =
    \fsym(c^{m+n_2}[s],c^{m+n_3}[s]) = \fsym(m+n_2, m+n_3)$.
  \end{itemize}
\end{proof}

We consider the following polynomials in the indeterminate
$i \in \nat$. We define them in a mutually recursive way,
which reflects the mutually recursive nature of $r_1$ and
$r_2$ and hence facilitates the proof of the existence of
a $(r_1,r_2,\rra_P)$-chain (Prop.~\ref{prop:binary-chain}
below).
\begin{definition}
  We let
  \begin{itemize}
    \item $\Pi_0(i) = n_2$ and $\Pi'_0(i) = n_3$
    \item $\Pi_{n+1}(i) = \Delta_n(i) + n_2$ and
    $\Pi'_{n+1}(i) = \Delta'_n(i) + n_3$
    for all $n \in \nat$
  \end{itemize}
  where, for all $n \in \nat$,
  \begin{itemize}
    \item $\Delta_n(i) = 0$ if $t = s$ and
    $\Delta_n(i) = \Delta'_n(i)$ if $t = x$
    \item $\Delta'_n(i) = i\Pi'_n(i) + \Pi_n(i)$.
  \end{itemize}
\end{definition}

\begin{example}\label{ex:binchain-lp-2}
  In Ex.~\ref{ex:binchain-lp}, we have
  $t = x$ and $(n_1,n_2,n_3) = (1,0,1)$.
  Hence:
  \begin{itemize}
    \item $\Pi_0(i) = n_2 = 0$
    and $\Pi'_0(i) = n_3 = 1$
    \item $\Pi_1(i) = \Delta_0(i) + n_2 =
    \Delta'_0(i) = i\Pi'_0(i) + \Pi_0(i) = i$
    \item $\Pi'_1(i) = \Delta'_0(i) + n_3 = i + 1$
    \item $\Pi_2(i) = \Delta_1(i) + n_2 =
    \Delta'_1(i) = i\Pi'_1(i) + \Pi_1(i) =
    i^2 + i + i = i^2 + 2i$
    \item $\Pi'_2(i) = \Delta'_1(i) + n_3 = i^2 + 2i + 1$
  \end{itemize}
\end{example}

The next lemma provides a simpler form of $\Pi$ and $\Pi'$
for the case $t = s$ (the case $t = x$ is more intricate).
\begin{lemma}\label{lem:poly-t-is-s}
  If $t = s$ then, for all $n \in \nat$,
  $\Pi_n(i) = n_2$ and
  $\Pi'_n(i) = n_3i^n + \sum_{k=0}^{n-1} (n_2+n_3)i^k$.
\end{lemma}
\begin{proof}
  Suppose that $t = s$. Then, for all $n \in \nat$,
  $\Delta_n(i) = 0$, so $\Pi_{n+1}(i) = n_2$. As
  $\Pi_0(i) = n_2$ also, for all $n \in \nat$
  we have $\Pi_n(i) = n_2$. Now, we prove that
  $\Pi'_n(i) = n_3i^n + \sum_{k=0}^{n-1} (n_2+n_3)i^k$.
  We proceed by induction on $n$.
  \begin{itemize}
    \item (Base: $n = 0$) We have
    $\Pi'_n(i) = n_3 = n_3i^n + \sum_{k=0}^{n-1} (n_2+n_3)i^k$.
    \item (Induction) Suppose that the property holds for
    some $n \in \nat$. We have
    $\Pi'_{n+1}(i) = \Delta'_n(i) + n_3 = i\Pi'_n(i) + \Pi_n(i) + n_3$.
    But, as $t = s$, $\Pi_n(i) = n_2$ and, by induction
    hypothesis, $\Pi'_n(i) = n_3i^n + \sum_{k=0}^{n-1} (n_2+n_3)i^k$.
    So, $\Pi'_{n+1}(i) = i(n_3i^n +
    \sum_{k=0}^{n-1} (n_2+n_3)i^k) + n_2 + n_3 =
    n_3i^{n+1} + \sum_{k=0}^n (n_2+n_3)i^k$.
  \end{itemize}
\end{proof}

\begin{example}\label{ex:binchain-trs-2}
  In Ex.~\ref{ex:binchain-trs}, we have
  $t = s$ and $(n_1,n_2,n_3) = (2,1,0)$.
  Hence, by Lem.~\ref{lem:poly-t-is-s},
  we have $\Pi_n(i) = 1$
  and $\Pi'_n(i) = \sum_{k=0}^{n-1} i^k$ 
  for all $n \in \nat$.
\end{example}

Using $\Pi$ and $\Pi'$, we define the set of terms $A$:
\begin{definition}
  We let $A = \{a_n = \fsym(\Pi_n(n_1), \Pi'_n(n_1))
  \mid n \in \nat\}$.
\end{definition}

Now we prove the existence of the $(r_1,r_2,\rra_P)$-chain
\[a_0 \mathop{(\rra^{\Pi'_0(n_1)}_{r_1} \circ \rra_{r_2})}
a_1 \mathop{(\rra^{\Pi'_1(n_1)}_{r_1} \circ \rra_{r_2})}
a_2 \mathop{(\rra^{\Pi'_2(n_1)}_{r_1} \circ \rra_{r_2})} \cdots\]
\begin{proposition}\label{prop:binary-chain}
  For all $n \in \nat$, we have
  $a_n \mathop{(\rra_{r_1}^{\Pi'_n(n_1)} \circ \rra_{r_2})} a_{n+1}$.  
\end{proposition}
\begin{proof}
  Let $n \in \nat$. We have $a_n = \fsym(\Pi_n(n_1),\Pi'_n(n_1))$.
  By Lem.~\ref{lem:reductions-r1} and Lem.~\ref{lem:reductions-r2}, 
  \[a_n \rra_{r_1}^{\Pi'_n(n_1)}
  \fsym\big(\underbrace{n_1 \times \Pi'_n(n_1) + \Pi_n(n_1)}_{\Delta'_n(n_1)}, 0\big)
  \rra_{r_2}
  \fsym\big(m, \underbrace{\Delta'_n(n_1) + n_3}_{\Pi'_{n+1}(n_1)}\big)
  \]
  where $m = n_2 = \Pi_{n+1}(n_1)$ if $t = s$ and
  $m = \Delta'_n(n_1) + n_2 = \Pi_{n+1}(n_1)$ if $t = x$.
  Hence, $a_n \mathop{(\rra_{r_1}^{\Pi'_n(n_1)} \circ \rra_{r_2})} a_{n+1}$.
\end{proof}

\begin{example}
  In Ex.~\ref{ex:binchain-trs}, we have 
  $\Pi_n(i) = 1$ and $\Pi'_n(i) = \sum_{k=0}^{n-1} i^k$ 
  for all $n \in \nat$ (see Ex.~\ref{ex:binchain-trs-2}).
  We also have $n_1 = 2$ and the $(r_1,r_2,\rra_P)$-chain:
  \[\underbrace{\fsym(\ssym(\zero), \zero)}_{a_0}
  \rra^{\Pi'_0(n_1)}_{r_1}
  \fsym(\ssym(\zero), \zero)
  \rra_{r_2}
  \underbrace{\fsym(\ssym(\zero), \ssym(\zero))}_{a_1}
  \rra^{\Pi'_1(n_1)}_{r_1}
  \fsym(\ssym^3(\zero), \zero)
  \rra_{r_2}
  \underbrace{\fsym(\ssym(\zero), \ssym^3(\zero))}_{a_2}
  \rra^{\Pi'_2(n_1)}_{r_1}
  \cdots\]
\end{example}

\begin{example}
  In Ex.~\ref{ex:binchain-lp}, we have
  $\Pi_0(n_1) = 0$, $\Pi'_0(n_1) = 1$,
  $\Pi_1(n_1) = 1$, $\Pi'_1(n_1) = 2$,
  $\Pi_2(n_1) = 3$, $\Pi'_2(i) = 4$, \dots{}
  (see Ex.~\ref{ex:binchain-lp-2}).
  We have the $(r_1,r_2,\rra_P)$-chain:
  \[\underbrace{\fsym(\zero, \ssym(\zero))}_{a_0}
   \rra^{\Pi'_0(n_1)}_{r_1}
  \fsym(\ssym(\zero), \zero)
  \rra_{r_2} 
  \underbrace{\fsym(\ssym(\zero), \ssym^2(\zero))}_{a_1}
  \rra^{\Pi'_1(n_1)}_{r_1}
  \fsym(\ssym^3(\zero), \zero)
  \rra_{r_2}
  \underbrace{\fsym(\ssym^3(\zero), \ssym^4(\zero))}_{a_2}
  \rra^{\Pi'_2(n_1)}_{r_1}
  \cdots\]
\end{example}

\section{Future Work and Implementation}
We plan to investigate how our work relates to the forms
of non-termination detected by the approaches
of~\cite{emmesEG12,geserZ99,wangS06}.
We have no clear idea for the moment.

Our tool \nti{} (Non-Termination Inference)~\cite{nti}
is designed to automatically prove the existence of
infinite chains in TRSs and in LPs. It first transforms
the original program $P$ into a program $P'$: for TRSs,
it uses the dependency pairs combined with a variant of
the overlap closure~\cite{payet18} and, for LPs, it uses
the binary unfolding~\cite{codishT99,gabbrielliG94}.
By~\cite{artsG00,codishT99,guttagKM83}, non-termination
of $P'$ implies that of $P$. Then, it detects 
recurrent pairs (Def.~\ref{def:rec-pair}), hence
binary chains (Prop.~\ref{prop:binary-chain}), in $P'$.

\bibliographystyle{plainurl}

\end{document}